\documentstyle[twocolumn,aps,epsf]{revtex}           %   <---| galley
\begin{document}                          	     %       |
\draft                                    	     %       |
\twocolumn[\hsize\textwidth\columnwidth\hsize\csname %	     | 
@twocolumnfalse\endcsname

\title{Crossover Scaling in Dendritic Evolution at Low Undercooling\\} 

\author{Nikolas Provatas$^{1,2}$, Nigel Goldenfeld$^{1}$, 
and Jonathan Dantzig$^{2}$ }

\address{
$^1$University of Illinois at Urbana-Champaign, Department 
of Physics \\1110 West Green Street, Urbana, IL, 61801 
}
\address{
$^2$  University of Illinois at Urbana-Champaign, Department
of Mechanical and Industrial Engineering
\\1206 West Green Street, Urbana, IL, 61801 
}

\author{Jeffrey C.~ LaCombe$^{3}$, Afina Lupulescu$^{3}$, 
Matthew B.~Koss$^{3}$, and Martin E.~Glicksman$^{3}$}

\address{
$^3$ Rensselaer Polytechnic Institute \\110 $8^{th}$ Street, Troy, New York 12180-3590
}
\author{Robert Almgren$^{4}$}
\address{
$^4$ University of Chicago, Department of Mathematics, Chicago, IL 60637
}

\date{\today}

\maketitle

\begin{abstract}
We examine scaling in two-dimensional simulations of dendritic growth 
at low undercooling, as well as in three-dimensional pivalic acid dendrites 
grown on NASA's USMP-4 Isothermal Dendritic Growth Experiment. We report
new results on self-affine evolution in both the experiments and 
simulations. We find that the time dependent scaling of our low undercooling 
simulations displays a cross-over scaling from a regime 
different than that characterizing Laplacian growth to steady-state growth.

\end{abstract}

% abstract is 525 characters 
\pacs{05.70.Ln, 81.30.Fb, 64.70.Dv, 81.10.Aj }
]

Recent computational and experimental advances in dendritic
growth offer a realistic prospect for a first principles
understanding of solidification microstructure formation.
Early experiments \cite{Hua81,Gli84} 
by Glicksman and coworkers on succinonitrile (SCN) provided the 
first benchmarks for theoretical models of dendritic growth. 
Comparison of experiments with theory has been difficult, 
however, since experiments were influenced by convection effects and 
performed at low undercooling using materials with low anisotropy, 
parameters for which computation is difficult.  
Such calculations can nevertheless be performed
in two-dimensions (2D) with state-of-the-art numerical methods 
combining so-called phase-field models 
\cite{Lan86a,Cag86a,Cag86b,Col85,War95,Whe92,Eld94,Whe96,Kob93,Pro96,Wan96}
and adaptive-mesh refinement\cite{Bra97,Pro98a,Pro98b}.
In the most recent round of experiments\cite{Mic98,IDGE_des} 
Glicksman and coworkers have reported observations on Pivalic 
Acid (PVA), whose higher anisotropy brings the benchmarks closer 
to the parameter range of theoretical computations.  

Predicting dendritic growth theoretically
has focused on the tip speed and shape in the steady state.  
simulations in 2D by Karma\cite{Kar95} and subsequently
ourselves\cite{Pro98a} have convincingly shown that the dynamically
selected steady state is indeed the fastest of the discrete set of
allowed needle crystals, as predicted by solvability
theory\cite{Ben84,Kes84,Kes88,Bre91,Pom91}.  However, at low
undercoolings the diffusion length is so large that the time needed for
each dendritic arm of a growing crystal to be in isolation from the
others becomes much longer than any realistic simulation time \cite{Pro98a}.  
This regime, where dendrite arms cannot simply translate at a uniform speed
because of their mutual interactions, was first systematically analyzed
by Almgren et al. \cite{Alm93b}, who used solvability theory to explore
the case when the temperature field strictly obeys Laplace's equation,
with constant flux, at each time.  They demonstrated that the
dendrite tip position grows with the 3/5 power of time, whereas the width
grows with the 2/5 power, results which were consistent with subsequent
experiments in Hele-Shaw flow\cite{Ign96}.

In this Letter we explore dendritic growth 
dynamics at low undercooling, using the full diffusion equation dynamics. 
We find that the time-dependent evolution of 2D dendrite profiles 
is self-affine in time, generalizing the results of Ref. \cite{Alm93a}
for the case of growth with a non-constant flux. Underlying this 
scaling behavior is a power law dependence on time of the dendrite 
tip position and maximum dendrite width. We find that scaling of these 
quantities displays a cross-over from a growth regime different 
from that of Hele-Shaw flow, to one characterized by steady-state 
tip growth. Meanwhile, comparison of our low undercooling simulations 
with microscopic solvability theory gives good agreement for the value of
the so-called stability parameter.  We also examine scaling in 3D dendrite 
data on pivalic acid obtained from NASA's USMP-4 Isothermal Dendritic 
Growth Experiment (IDGE), also finding self-affine scaling in the global 
time-dependent PVA dendrite profiles. 

The simulated dendrites are modeled using the phase-field model employed
in \cite{Kar95}. Temperature $T$ is rescaled to
$U=c_P(T-T_M)/L$, where $c_P$ is the specific heat at constant pressure,
$L$ is the latent heat of fusion and $T_M$ is the melting temperature. 
The order parameter is defined by $\phi$, with $\phi=1$ in the solid,
and $\phi=-1$ in the liquid, and the interface defined by $\phi=0$. In what 
follows time is rescaled by the time scale $\tau_o$ characterizing atomic 
movement in the interface, and length by the length scale $W_o$ characterizing the  
width of the liquid--solid interface. The model is given by
\begin{eqnarray}
&&\frac{\partial U}{dt} = D \nabla^2 U + \frac{1}{2} \frac{\partial 
\phi}{\partial t}
\label{phase-field}\\
\nonumber
&A^2(\vec{n})&  \frac{\partial \phi}{dt} = \nabla \cdot 
(A^2(\vec{n}) \nabla \phi )  +  
(\phi - \lambda U (1-\phi^2))(1- \phi^2) \\
\nonumber
& + &
\frac{\partial }{\partial x} \left( |\nabla \phi|^2 A(\vec{n}) 
\frac{\partial A(\vec{n})}{\partial \phi_{,x}} \right) +
\frac{\partial }{\partial y} \left( |\nabla \phi|^2 A(\vec{n}) 
\frac{\partial A(\vec{n})}{\partial \phi_{,y}} \right),
\end{eqnarray}
where $D=\alpha \tau_o/W_o^2$, $\alpha$ is the thermal diffusivity,
and $\lambda$ controls the coupling of $U$ and $\phi$.  Anisotropy has been
introduced in Eqs.~(\ref{phase-field}) by defining the width of the 
interface to be $W(\vec{n})=W_o A(\vec{n})$ and the characteristic 
time by $\tau(\vec{n})=\tau_o A^2(\vec{n})$ \cite{Kar95}, where 
$A(\vec{n}) \in [0,1]$, and 
%
%\begin{equation}
$
A(\vec{n}) = (1- 3 \epsilon) \left[  1 + \frac{4 \epsilon }{ 1 - 3 \epsilon}
\frac{(\phi_{,x})^4 + (\phi_{,y})^4 }{| \nabla \phi|^4}\right].
$
%\label{width}
%\end{equation}
%
The vector 
$\vec{n}=(\phi_{,x}\hat{x}+\phi_{,y}\hat{y})/(\phi_{,x}^2+\phi_{,y}^2)^{1/2}$ 
is the normal to the contours of $\phi$, 
and $\phi_{,x}$ and $\phi_{,y}$ represent partial derivatives with 
respect to $x$ and $y$.  The constant $\epsilon$ parameterizes the 
deviation of $W(\vec{n})$ from $W_o$. We expect the results to be similar for
other definitions of anisotropy  \cite{Cag87}.
The parameters of Eqs.~(\ref{phase-field}) are related to the
appropriate Stefan problem using the relationships given in 
\cite{Kar95}. In particular, $W$, $\tau$, $\lambda$ and $D$ may 
be chosen to simulate an arbitrary, anisotropic capillary length 
$d(\vec{n})$, and interface attachment coefficient $\beta(\vec{n})$, 
which we chosen here as $\beta=0$, a limit appropriate for SCN and PVA.

Simulated dendrites were computed by solving Eqs.~(\ref{phase-field}) 
using the adaptive-grid method of Ref. \cite{Pro98a,Pro98b}. 
Simulated dendrites were grown in a 2D quarter-infinite space 
using zero-flux boundary conditions along the sides of the system.
Growth was initiated by a small quarter disk of radius $R_o$ centered at the 
origin. The preferred growth directions are 
along the x and y axes, making these the directions of growth of dendrite branches.
The order parameter is initially set to its equilibrium value 
$\phi_o(\vec{x})=-\tanh((|\vec{x}|-R_o)/\sqrt{2} )$ 
along the interface.  The initial temperature decays exponentially from $U=0$ at the
interface to its far-field, undercooled value $-\Delta$ as $\vec{x} \rightarrow \infty$.
Simulation data presented in this paper were obtained for three
undercoolings: $\Delta=0.25$ and $0.1$ and $0.05$. Details of these data are
presented in Table 1. The two data sets for $\Delta=0.1$ correspond to 
different minimum grid spacings $\Delta x_{\rm min}$ \cite{Pro98a,Pro98b}.
Seed radii used in our simulations were $R_o=8.5, 15, 30, 30$ for 
$\Delta=0.25, 0.1(A), 0.1(B), 0.05$,
respectively. In all cases $R_o$ is smaller than the thermal diffusion 
length by a factor of 20 or greater.

The results of our low undercooling simulations are also 
contrasted here with new experimental data obtained from PVA dendrites.  These
experiments were performed by four of the authors (LaCombe, Lupulescu,
Koss and Glicksman) during NASA's USMP-4 Isothermal Dendritic Growth
Experiment (IDGE).  This experiment is described in detail elsewhere
\cite{IDGE_des}.  The IDGE experiment is designed to study dendrites grown
under microgravity conditions, where transport in this particular
process is considered to be conduction-limited.  The crystals are grown
in an undercooled melt, controlled to within 0.001K. 
Growth is monitored thermometrically, while images are obtained from
two perpendicular directions using video and still cameras (electronic
and film).  Experimental results presented here were compared with four
independent data subsets for dendrites grown at undercoolings of 0.58K,
0.63K, and 0.47K.  We present the results from experiments corresponding
to $\Delta = 0.052$.  These data were captured at times $t_1=42.48$, $t_2=62.7$3, 
and $t_3=82.98$ seconds after the dendrite was detected.  The anisotropy for PVA was
estimated at $\epsilon_{\rm pva} = 0.025$ \cite{Mus92}.

We found the individual primary arms of our simulated dendrites to 
be self-affine, beyond some transient time, at  all undercoolings 
examined. Figure~1 shows the ($\Delta$-dependent) scaling profile 
for 2D dendrites grown at $\Delta=0.05$ and $\Delta=0.25$, respectively. 
The global scaling profile is obtained by scaling the x-direction by 
$(x-x_b)/X_{\rm max}$, where $X_{\rm max}(t)$ is the distance from the 
tip $x_{\rm tip}(t)$ to the base $x_b(t)$ of the dendrite 
arm, and  the y-direction by $y/Y_{\rm max}(t)$, the maximum half-width of the 
lateral dimension of the primary dendrite arm.  The tip and transverse directions
were found to scale as $X_{\rm max} \sim t^{\beta}$ and $Y_{\rm max} \sim t^{\gamma}$,
where for $\Delta=0.05,0.1A, 0.1B$ and $0.25$, $\beta=0.0.73, 0.73, 0.78, 0.97$, 
and $\gamma=0.43, 0.43, 0.45$, and $0.55$, respectively.  For the $\Delta=0.25$ 
data, which at late times contained sidebranching induced by lattice-noise, 
we define $Y_{\rm max}(t)$ using the {\it mean } interface position, obtained 
by smoothing the data. This definition of the sidebranch envelope gives
different results than using the maximum of the sidebranch envelope \cite{Li97,Bil95}.

At low undercooling, long-lived transient interactions between 
neighboring primary dendrite arms causes their velocity and tip radius
to deviate (within simulation time scales) from
their steady-state values predicted by solvability 
theory \cite{Pro98a}. However, we do find that the {\it stability 
parameter} $\sigma^*=2d_o D/VR^2$, where $V$ and $R$ are the time-dependent 
velocity and tip radius, agrees well with the value predicted by 
solvability theory.  Figure~2 shows $\sigma^*$ vs. time from our 
simulations at $\Delta=0.25$, $0.1$ and $0.05$.  Error bars were estimated using
$\Delta V$, the fluctuations in velocity, and $\Delta R$, deviations in 
radius of curvature. The radius was obtained by fitting to a second 
order polynomial near the tip. Deviations in the fit gave an estimate for $\Delta R$. 
Data for $\Delta=0.1$ set B, omitted for clarity, converge to approximately the same 
$\sigma^*$ as the $\Delta=0.1$ set A data, but display somewhat larger 
fluctuations around the mean, due to the larger grid spacing used.

The time-dependent behavior of the tip position and lateral growth rate
of our 2D dendrites are characterized by the scaling of $X_{\rm max}(t)$ and 
$Y_{\rm max}(t)$. Figure~3 shows $X_{\rm max}$ and $Y_{\rm max}$ scaled onto  
respective crossover functions of the form
\begin{eqnarray}
X_{\rm max}(t)/L_D = \frac{t}{\tau_D} F_X(t/\tau_D),
\label{crossoverx}
\end{eqnarray}
and
\begin{eqnarray}
Y_{\rm max}(t)/L_D = (\frac{t}{\tau_D})^{1/2} F_Y(t/\tau_D)
\label{crossovery}
\end{eqnarray}
The parameters $L_D$ and $\tau_D$ are effective diffusion length and time scales
characterizing the intermediate regime,  and are fit to give collapse of 
the $X_{\rm max}$ and $Y_{\rm max}$ data. 
The data for $F_X(z)$ show a crossover scaling from fit to approximately 
$F_X(z) \sim z^{-0.25}$ at early times to $F_X(z) \sim z^{-0.03}$ 
in the steady-state regime. The cross-over in $F_Y(z)$ is given by 
$F_Y(z) \sim z^{-0.07}$ at small $z$ to $F_Y(z) \sim z^{0.05}$ at large 
arguments of $F_Y(z)$.  Exponent errors were approximately $\pm 0.02$,
except for the $\Delta=0.25$ data at late time, where they were
$\pm 0.05$.  These asymptotic limits are demonstrated by the leveling off
of $F_X(\chi)$ and $F_Y(\chi)$ as $\chi=t/\tau_D$ becomes large.  

Our simulations are in a regime where the dendrite is
much smaller than the diffusion length. However, we do not observe the 
early-time scaling described by Almgren, et al  \cite{Alm93b}
since their calculations assume that dendrites
are grown with a {\em constant} flux, whereas in our simulations the 
far field is diffusive with a specified small undercooling. 
To illustrate this difference, let us assume that the rate of change 
of solid-fraction evolves as $F \sim t^{f}$, whereby the solidified area 
$A \sim t^{1+f}$. Since $X_{\rm max} \sim t^{\beta}$, 
$Y_{\rm max} \sim t^{\gamma}$, $A \sim X_{\rm max} Y_{\rm max}$, and so
$1+f = \beta + \gamma$. Since $VR^2 =const$, we obtain the scaling 
relation $S(f,\beta)=4f -5\beta +3=0$. Our early-time exponents give
$S(0.18,0.75)=-0.03 \pm 0.1$. In the late-time regime, 
$S(0.52,0.97)=0.23 \pm 0.22$. The late-time error in $S$ arises when 
we estimate $\gamma$ by smoothing the dendrite profiles for $\Delta=0.25$ 
($\epsilon=0.05$).  For comparison, we produced data for $\Delta =0.25$, 
$\epsilon=0.025$ (not in Table 1), and which were free of spurious sidebranches, 
obtaining $Y_{\rm max} \sim t^{0.5}$.

Self-affine time-dependent scaling was also found in the
mean dendrite profiles of the new 3D IDGE PVA data.
Figure~4 shows the scaled PVA data for $t=t_1, t_2, t_3$. For
comparison these data are superimposed on our 2D simulation data
for $\Delta=0.05$, $\epsilon=0.025$.  There is a slight asymmetry
in the PVA data, likely due to interactions with other dendrite arms.
For this reason, we scaled with respect to the top side the dendrite
profile. Similar scaling was observed in all four
IDGE data sets. The experimental and simulated profiles show clear
differences near the tip, as one would expect. Curiously, however, 
the profile shapes are in good agreement away from the tip. 
Similar results were also found in our $\Delta=0.1$ data.
The reason for this is illustrated in Fig.~1, which shows that 
the difference in 2D dendrite profiles is small away from the tip.

As plausible explanation for the apparent agreement between our
low undercooling 2D simulations and our 3D experimental
data, we note that away from the tip, the diffusion field is
more cylindrically symmetric than at the tip because the local
diffusion length is larger. Thus diffusion of heat away from
the interface is better approximated by the 2D diffusion equation. 
We hope to examine this idea critically, as well as to accurately 
determine the experimental scaling behavior in future publications.

%%%%%%%%%%%%%%%%%%%%%%%%%%%%%%%%%%%%%%%%%%%%%%%%%%%%%%%%%%%%%%
We thank Wouter-Jan Rappel for the code to obtain solvability results, 
Terry Chay for useful discussions.
We also thank Julie Frei, and Douglas Corrigan for assistance 
in obtaining the PVA dendrite profiles.  This has been supported 
by the NASA Microgravity Research Program, under Grants NAG8-1249, 
and NAS3-25368.
%%%%%%%%%%%%%%%%%%%%%%%%%%%%%%%%%%%%%%%%%%%%%%%%%%%%%%%%%%%%%%

%%%%%%%%%%%%%%%%%%%%%%%%%%%%%%%%%%%%%%%%%%%%%%%%%%%%
%BIBLIOGRAPHY
%%%%%%%%%%%%%%%%%%%%%%%%%%%%%%%%%%%%%%%%%%%%%%%%%%%%%%%%

%\bibliography{biblio}

%%%%%%%%%%%%%%%%%%%%%%%%%%%%%%%%%%%%%%%%%%%%%%%%%%%%
%TABLES AND FIGURE CAPTIONS 
%%%%%%%%%%%%%%%%%%%%%%%%%%%%%%%%%%%%%%%%%%%%%%%%%%%%%%%%

\begin{table}[h] \caption{\label{tab1} Parameters for simulated dendrites. The time $t^*=255622.4$  }
\begin{center}
\begin{tabular}[t]{|c|c|c|c|c|c|c|c|}\hline
$\Delta$ & $\epsilon$ & $\Delta x_{\rm min}$ & $\Delta t$ & D  & $d_o$ & $L_x$ & $L_y$\\
\hline
0.25     & 0.05       & 0.78           & 0.048      & 13 & 0.043   & 12800  & 6400  \\
0.1(A)   & 0.05       & 0.78           & 0.08       & 13 & 0.043   & 102400 & 51200 \\
0.1(B)   & 0.05       & 1.56           & 0.08       & 30 & 0.01846 & 102400 & 51200 \\
0.05($t<t^*$)         & 0.025          & 1.56       & 0.03       & 40 & 0.01385   & 102400 & 51200 \\
0.05($t>t^*$)         & 0.025         & 0.78        & 0.03       & 40 & 0.01385   & 102400 & 51200 \\
\hline
\end{tabular}
\end{center}
\end{table}

\begin{figure}[h]
\begin{center}
\leavevmode\mbox{\epsfxsize=3.0in\epsfbox{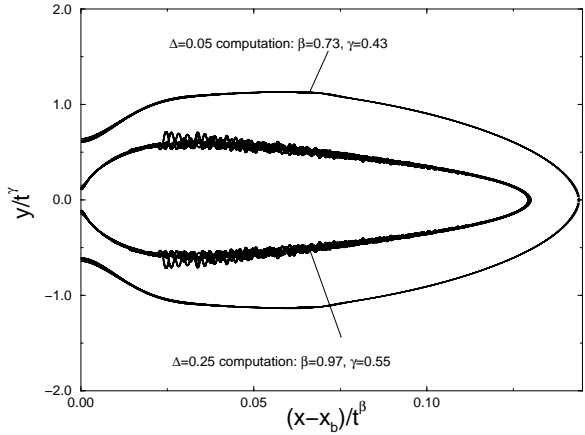}}
\end{center}
\caption{Comparison of scaled dendrite profiles for $\Delta=0.05$ and $\Delta=0.25$.
For $\Delta=0.25$, nine times are plotted, spaced between $28643 < t < 66083$.
For $\Delta =0.05$, there are six times in the range $222022 < t < 279622$.
}
\label{comp_profiles}
\end{figure}

\begin{figure}[h]
\begin{center}
\leavevmode\mbox{\epsfxsize=3.0in\epsfbox{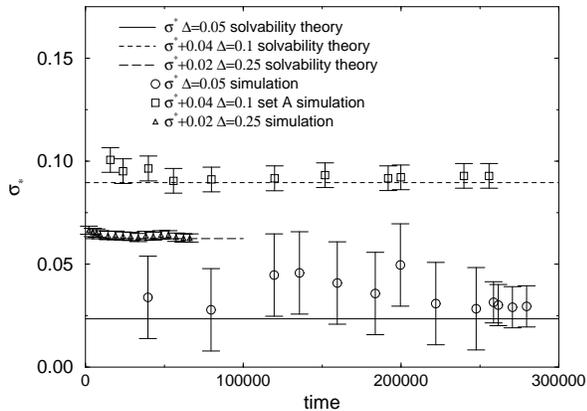}}
\end{center}
\caption{Simulation data of $\sigma^*$ vs. time for 
$\Delta=0.25$,$0.1$ (set A) and $0.05$. For clarity, the 
$\Delta=0.1$ and $0.25$ data have been shifted along the 
y-axis by $0.04$ and $0.02$, respectively.
}
\label{sigmas}
\end{figure}

\begin{figure}[h]
\begin{center}
\leavevmode\mbox{\epsfxsize=3.0in\epsfbox{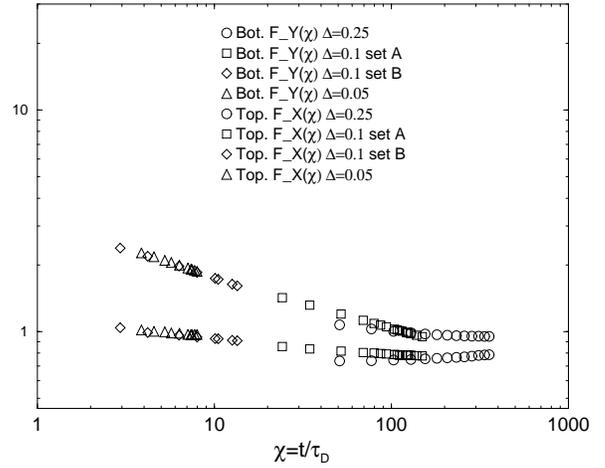}}
\end{center}
\caption{Crossover scaling functions describing lateral width of simulated dendrite arm 
$Y_{\rm max}$ and tip-to-base distance $X_{\rm max}$, 
for $\Delta=0.25$, $0.1$ (sets A and B)  and $0.05$. 
}
\label{scaling}
\end{figure}

\begin{figure}[h]
\begin{center}
\leavevmode\mbox{\epsfxsize=3.0in\epsfbox{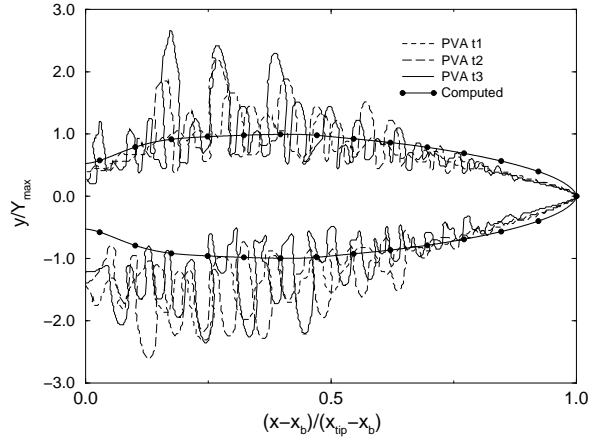}}
\end{center}
\caption{Scaled USMP-4 PVA dendrites grown at
$\Delta=0.052$. 2D simulation data is superimposed 
for $\Delta=0.05$, $\epsilon=0.025$.  Comparison of 2D and 3D 
data made merely to illustrate self-affinity in both 2D and 3D dendrites.
}
\label{DT0.05}
\end{figure}

%%%%%%%%%%%%%%%%%%%%%%%%%%%%%%%%%%%%%%%%%%%%%%%%%%%%%%%%%%%%%%%

\end{document}